\documentclass[aps,pre,reprint,amsmath,amssymb,superscriptaddress]{revtex4-2}

\usepackage{graphicx}
\usepackage{dcolumn}
\usepackage{bm}
\usepackage{amsmath}
\usepackage{amsthm}
\usepackage{amssymb}
\usepackage{cases}
\usepackage{url}

\begin{document}


\title{A Forced Harmonic Oscillator, Interpreted as Diffraction of Light}


\author{Toshihiko Hiraiwa}
\email{hiraiwa@spring8.or.jp}
\affiliation{RIKEN SPring-8 Center (RSC), Sayo, Hyogo, 679-5148, Japan}
\author{Kouichi Soutome}
\affiliation{RIKEN SPring-8 Center (RSC), Sayo, Hyogo, 679-5148, Japan}
\affiliation{Japan Synchrotron Radiation Research Institute (JASRI), Sayo, Hyogo, 679-5198, Japan}
\author{Hitoshi Tanaka}
\affiliation{RIKEN SPring-8 Center (RSC), Sayo, Hyogo, 679-5148, Japan}


\date{\today}

\begin{abstract}
We investigate a simple forced harmonic oscillator with a natural frequency varying with time. It is shown that the time evolution of such a system can be written in a simplified form
with Fresnel integrals, as long as the variation of the natural frequency is sufficiently slow compared to the time period of oscillation. Thanks to such a simple formulation, we found, for the first time, that
a forced harmonic oscillator with a slowly-varying natural frequency is essentially equivalent to diffraction of light.
\end{abstract}


\maketitle

\section{Introduction}
Resonance phenomena, in conjunction with forced harmonic oscillators~(FHOs), are observed in a lot of dynamical systems, and are discussed as a fundamental problem in
standard textbooks on classical mechanics~\cite{feynman,goldstein}. The concept of resonances is present in many branches of science, and therefore has a wide variety of
applications. About three hundred years after the discovery of a resonance-like phenomenon, theoretical models for FHOs with characteristic resonances have been well
established~(see Refs.~\cite{nature_res,review} for a recent historical review of FHOs).  In many cases, FHOs have been discussed in the context of resonance phenomena. Here, we present a
simple formulation of a FHO with a natural frequency varying with time using Fresnel integrals~\cite{fresnel}. Thanks to such a simple formulation, we found, for the first time, that a FHO
with a time-varying natural frequency is essentially equivalent to diffraction of light from a single slit, i.e., so-called Fraunhofer or Fresnel diffraction~\cite{sommerfeld,optics_1,optics_2,optics_3}.

\section{Formulation}
In this article, we investigate a simple FHO with a time-varying natural frequency. We suppose that the driving force is activated at $t = 0$ and is then deactivated at $t = \Delta \; (>0)$,
and that the frequency of the driving force~($\omega_{f} \equiv 2\pi\nu_{f}$) is kept constant while the natural frequency of the oscillator~($\omega \equiv 2 \pi \nu$) varies with time as
$\omega(t=0) < \omega_{f} < \omega(t = \Delta)$.
In addition, it is assumed that $\omega(t)$ varies very slowly compared to the time period of oscillation, namely:
\begin{equation}
\label{eq:assump}
|\dot{\omega}(t)| \ll \omega^{2}(t), ~~~~ |\ddot{\omega}(t)| \ll \omega^{3}(t),
\end{equation}
\noindent
where $\dot{\omega}(t)$ and $\ddot{\omega}(t)$ represent the first and second derivatives of $\omega(t)$, respectively. 

The basic equation of motion for the above system is written in the form:
\begin{equation}
\label{eq:basic_eq}
\ddot{x} + \omega^{2}(t)x = F(t),
\end{equation}
\noindent
with the driving force:
\begin{equation}
F(t) =
\begin{cases}
0 & (t < 0) \\
F_{0}\cos\left( \omega_{f}t + \phi_{0} \right) & (0 \leq t \leq \Delta) \\
0 & (t > \Delta)
\end{cases},
\end{equation}
\noindent
where $x$ denotes displacement from the equilibrium position as a function of $t$, $F_{0}$ is the amplitude of the sinusoidal force, and $\phi_{0}$ is a constant phase. Here, we neglect a damping
term for simplicity~\footnote{The same discussion can be made even when a damping term is included, as long as its effect is sufficiently weak.}.

Now, the frequency $\omega $~($= 2\pi \nu$) of the oscillator is a function of $t$, and can be expanded in a Taylor series:
\begin{equation}
\label{eq:expand_freq}
\begin{split}
\omega(t) &= \omega^{(0)} + \omega^{(1)}t + \frac{\omega^{(2)}}{2}t^{2} + \cdots \\
              &= 2\pi \left( \nu^{(0)} + \nu^{(1)}t + \frac{\nu^{(2)}}{2}t^{2} + \cdots \right). 
\end{split}
\end{equation}
\noindent
Here we adopt a linear approximation for Eq.~(\ref{eq:expand_freq}), namely:
\begin{equation}
\label{eq:linear_freq}
\begin{split}
\omega(t) &= \omega^{(0)} + \omega^{(1)}t \\
              &= 2\pi \left(\nu^{(0)} + \nu^{(1)}t \right).
\end{split}
\end{equation}
\noindent
It should be noted that this can be made without loss of generality because a linear approximation holds for an arbitrary function $\omega(t)$ as long as the time window $\Delta$ is taken to be
sufficiently short, i.e., $\Delta < 1/\omega^{(0)}$~[see Eq.~(\ref{eq:assump})]. For simplicity, we hereafter assume $\omega^{(1)} > 0$. Then the assumption~(\ref{eq:assump}) becomes:
\begin{equation}
\epsilon^{2} \equiv  \omega^{(1)}/\left( \omega^{(0)} \right)^{2} \ll 1.
\label{eq:assump_mod}
\end{equation}

Equation~(\ref{eq:basic_eq})  can be approximately solved with the aid of the well-known Green's Function method. Under the
assumption~(\ref{eq:assump})~[or (\ref{eq:assump_mod})], the Green's function of Eq.~(\ref{eq:basic_eq}) is given by~(see Appendix~\ref{app:green_fnc} for details): 
\begin{equation}
\label{eq:green_fnc}
G(t, t^{\prime}) = \frac{-i}{2\sqrt{\omega(t)\omega(t^{\prime})}}\exp\left[i\int^{t}_{t^{\prime}}\omega(\tau)d\tau \right] + \text{c.c.}.
\end{equation}

Using the Green's function of Eq.~(\ref{eq:green_fnc}) together with the assumption~(\ref{eq:assump_mod}), one can easily obtain a particular solution of Eq.~(\ref{eq:basic_eq}) for
$t >\Delta$~\footnote{Here, we are interested in a particular solution because it contains all the effects of the driving force.}:
\begin{align}
x(t) =& \displaystyle \int^{\Delta}_{0}G(t, t^{\prime})F_{0}\cos(\omega_{f}t^{\prime} + \phi_{0})dt^{\prime} \nonumber \\
     \simeq& \displaystyle  \frac{ i F_{0}}{4 \omega^{(0)}} e^{-i \displaystyle \int^{t}_{0} \omega(\tau)d\tau} \times h(t;r) + \text{c.c.},
     \label{eq:sol}
\end{align}
\noindent
with an envelope function $h(t;r)$:
\begin{equation}
h(t;r) = A(t) \times \tilde{h}(r),
\label{eq:envelop}
\end{equation}
\noindent
where $A(t)$ is a damping factor:\begin{equation}
A(t) = \sqrt{\omega^{(0)}/\omega(t)}=\sqrt{\frac{\omega^{(0)}}{\omega^{(0)}+\omega^{(1)}t}},
\label{eq:damp_fac}
\end{equation}
\noindent
and $\tilde{h}(r)$ is a response function:
\begin{align}
\tilde{h}(r) =& \displaystyle  \int^{\Delta}_{0} A(t^{\prime}) \left( \exp \left[ i \left\{  \omega^{(0)}(1 - r)t^{\prime} + \frac{\omega^{(1)}}{2}t^{\prime 2}  - \phi_{0} \right\} \right] \right. \nonumber \\
             &+ \displaystyle \left. \exp \left[ i \left\{  \omega^{(0)}(1 + r)t^{\prime} + \frac{\omega^{(1)}}{2}t^{\prime 2} + \phi_{0} \right\} \right] \right) dt^{\prime} \nonumber \\
             \simeq& \displaystyle  \int^{\Delta}_{0} A(t^{\prime})\exp \left[ i \left\{  \omega^{(0)}(1 - r)t^{\prime} + \frac{\omega^{(1)}}{2}t^{\prime 2} - \phi_{0} \right\} \right]dt^{\prime}. \label{eq:res_fnc}
\end{align}
\noindent
Here we define $r \equiv \omega_{f}/\omega^{(0)}$~($= \nu_{f}/\nu^{(0)}$), and, in deriving Eq.~(\ref{eq:res_fnc}), we neglect a rapidly-oscillating term~(i.e., the second term) in the integrand.
Note that a damping factor $A(t)$ originates from the natural frequency varying with time, not from the presence of the driving force~\cite{landau_mechanics}.

The response function of Eq.~(\ref{eq:res_fnc}) is further simplified: as we shall see later, for $t^{\prime} \gtrsim 1/\sqrt{\omega^{(1)}}$, there is almost no contribution to the integral 
because of rapid oscillation of the integrand. For $t^{\prime} \lesssim 1/\sqrt{\omega^{(1)}}$, on the other hand, the damping factor in the integrand can be written as $A(t^{\prime}) \sim 1 + \mathcal{O}(\epsilon)$.
Given a sufficiently small $\epsilon$, we have $A(t^{\prime}) \sim 1$ and hence:
\begin{equation}
\label{eq:res_fnc_simple}
\begin{split}
\tilde{h}(r) \simeq& \displaystyle  \int^{\Delta}_{0} \exp \left[ i \left\{  \omega^{(0)}(1 - r)t^{\prime} + \frac{\omega^{(1)}}{2}t^{\prime 2} - \phi_{0} \right\} \right]dt^{\prime} \\
      =& \frac{1}{\sqrt{2\nu^{(1)}}} \exp\left( -i \left[ \displaystyle \frac{\pi\left\{\nu^{(0)}(1-r)\right\}^{2}}{\nu^{(1)}} + \phi_{0} \right] \right) \\
      &\times  \displaystyle \int^{\sqrt{2\nu^{(1)}}\Delta}_{0} \exp\left[ i \frac{\pi}{2} \left\{ \tilde{t} + \displaystyle \frac{\sqrt{2}\nu^{(0)}(1-r)}{\sqrt{\nu^{(1)}}}  \right\}^{2}
       \right]d\tilde{t} \\
      =& \frac{1}{\sqrt{2\nu^{(1)}}} \exp\left( -i \left[ \displaystyle \frac{\pi \left\{\nu^{(0)}(1-r)\right\}^{2}}{\nu^{(1)}} + \phi_{0} \right] \right)  \\
      &\times  \left[ \left\{ C(u_{2}) - C(u_{1}) \right\}  + i \left\{ S(u_{2}) - S(u_{1}) \right\} \right],
\end{split}
\end{equation} 
\noindent
where $u_{1}$ and $u_{2}$ are given by:
\begin{equation}
\label{eq:width_ho}
\begin{cases}
u_{1} =& \displaystyle \frac{\sqrt{2}\nu^{(0)}(1 - r)}{\sqrt{\nu^{(1)}}} \\
u_{2} =& u_{1} + \sqrt{2\nu^{(1)}}\Delta
\end{cases},
\end{equation}
\noindent
and two functions $C(u)$ and $S(u)$ are so-called Fresnel integrals, defined as:
\begin{equation}
\label{eq:fresnel_int}
\begin{cases}
C(u) =& \displaystyle \int^{u}_{0}\cos\left( \displaystyle \frac{\pi}{2}v^{2} \right)dv \\
S(u) =& \displaystyle \int^{u}_{0}\sin\left( \displaystyle \frac{\pi}{2}v^{2} \right)dv
\end{cases}.
\end{equation}

As we see from Eq.~(\ref{eq:sol}), the particular solution obtained here is of a characteristic form: that is, the first part of the r.h.s. of Eq.~(\ref{eq:sol}) represents a propagating wave
with frequency modulation, whereas the last one represents a response of the oscillation amplitude to the frequency $\omega_{f}$ of the driving force. Furthermore, as we see in 
the response function of Eq.~(\ref{eq:res_fnc_simple}), the imaginary argument of the exponent in the integrand is a quadratic function of a variable $t^{\prime}$~($\tilde{t}$), thus yielding
Fresnel integrals.

Our formulation can be also extended to the other case where the frequency $\omega$ of the oscillator is kept constant while the frequency $\omega_{f}$ of the driving force varies slowly with time as 
$\omega_{f}(t=0) < \omega < \omega_{f}(t=\Delta)$, as discussed in Refs.~\cite{pass_res,pass_res2,pippard}. In this case, we have no damping factors, and a response function is a bit modified:
\begin{equation}
\label{eq:res_fnc_force}
\tilde{h}(\tilde{r}) = \displaystyle \int^{\Delta}_{0}\exp \left[ -i \left\{ \omega_{f}^{(0)}(1-\tilde{r})t^{\prime} + \frac{\omega_{f}^{(1)}}{2}t^{\prime 2}  + \phi_{0} \right\}  \right]dt^{\prime},
\end{equation}
\noindent
which yields Fresnel integrals as well. Here, we write the frequency $\omega_{f}$ as:
\begin{equation}
\label{eq:freq_force}
\omega_{f}(t) = \omega_{f}^{(0)} + \omega_{f}^{(1)}t + \frac{\omega_{f}^{(2)}}{2}t^{2} +  \cdots,
\end{equation}
\noindent
define $\tilde{r} \equiv \omega/\omega_{f}^{(0)}$, and neglect rapidly-oscillating terms in the integrand. Note that, strictly speaking, the assumption of the "slow change" of $\omega_{f}(t)$ is not necessary for the
derivation of Eq.~(\ref{eq:res_fnc_force}) because a Green's function can be obtained just by solving the equation of motion for a free HO with a constant natural frequency $\omega$~[cf. Eq.~(\ref{eq:homo_eq})].

\section{Analogy to diffraction of light}
\begin{figure}[t]
\includegraphics[clip,width=\linewidth]{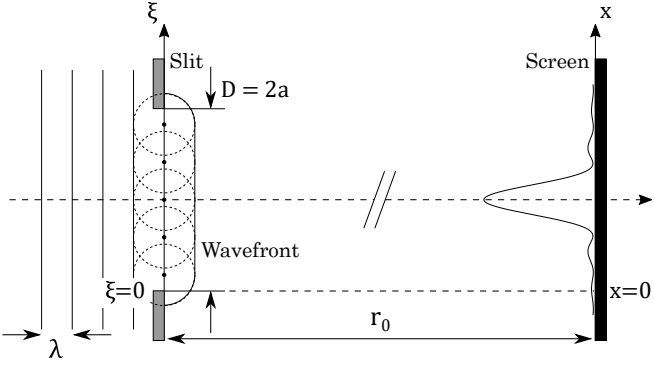}
\caption{\label{fig:diff_light}Diffraction of light from a single slit. Each solid line represents an undiffracted wavefront, while a solid curve represents a diffracted wavefront
determined from Huygens' wavelets at the aperture~(dashed-line circles). Here, we define the wavelength $\lambda$, the aperture size $D=2a$, and the distance $r_{0}$ between
the slit and the screen.}
\end{figure}

One may encounter a quite similar form as in Eqs.~(\ref{eq:sol}), (\ref{eq:res_fnc_simple}) and (\ref{eq:res_fnc_force}) in a description of diffraction of light from a single slit~(Fig.~\ref{fig:diff_light}) based on the so-called
Fresnel-Kirchhoff diffraction integral with the Fresnel approximation~(see, e.g., Ref.~\cite{optics_3}).  Fresnel's formulation of single-slit diffraction approximates the imaginary argument of the exponent in the integrand,
which represents a phase difference between secondary spherical waves from the wavefront at the aperture, to be a quadratic phase variation. Thus, the electric field on the screen, $E_{s}(x)$, can be written as:
\begin{equation}
E_{s}(x) = E_{0}e^{ikr_{0}} \displaystyle \int^{2a}_{0}\exp\left[ \frac{ik}{2r_{0}} \left(\xi - x \right)^{2} \right]d\xi,
\label{eq:diff_formula}
\end{equation}
\noindent
where $E_{0}$ is a constant field strength, and $k$ is a wave number~($=2\pi/\lambda$). In this case, we can also define an analogue function to Eq.~(\ref{eq:res_fnc_simple}):
\begin{equation}
\tilde{h}_{E}(x) = \displaystyle \int^{2a}_{0} \exp \left[ \frac{i\pi}{\lambda r_{0}} \left( \xi - x \right)^{2}  \right]d\xi.
\label{eq:res_func_diff}
\end{equation}

By comparing two functions $\tilde{h}$~[Eq.~(\ref{eq:res_fnc_simple})] and $\tilde{h}_{E}$~[Eq.~(\ref{eq:res_func_diff})], we can obtain exact relations that connect the two phenomena: to do so, we
introduce dimensionless integration variables, $\hat{t} \equiv t^{\prime}/\Delta$ and $\hat{\xi} \equiv \xi/(2a)$. Then we have the phase function of the integrand for Eq.~(\ref{eq:res_fnc_simple}):
\begin{equation}
\Phi(\hat{t}) = 2\pi\nu^{(0)}\Delta(1-r)\hat{t} + 2\pi\nu^{(1)}\Delta^{2}\frac{\hat{t}^{2}}{2} - \phi_{0}
\label{eq:phase_FHO}
\end{equation}
\noindent
and that for Eq.~(\ref{eq:res_func_diff}):
\begin{equation}
\Phi_{E}(\hat{\xi}) = -4\pi N_{F}\frac{x}{a}\hat{\xi} + 8\pi N_{F} \frac{\hat{\xi}^{2}}{2} + \pi N_{F}\left(\frac{x}{a}\right)^{2},
\label{eq:phase_diff}
\end{equation}
\noindent
where $N_{F}$ is a so-called Fresnel number:
\begin{equation}
N_{F} = \frac{a^{2}}{\lambda r_{0}}.
\label{eq:fres_num_light}
\end{equation}
\noindent
Since Eqs.~(\ref{eq:phase_FHO}) and (\ref{eq:phase_diff}) are both functions of a dimensionless variable, one immediately obtains the following relations:
\begin{align}
\nu^{(0)}\Delta(r-1) &\Longleftrightarrow 2N_{F}\frac{x}{a}, \label{eq:corresp_1}\\
\nu^{(1)}\Delta^{2} &\Longleftrightarrow 4N_{F}. \label{eq:corresp_2}
\end{align}

In the theory of single-slit diffraction, a Fresnel number $N_{F}$ is often defined to characterize diffraction patterns with different configurations: for $N_{F} \ll 1$, where the screen is
far from the slit, or where the slit aperture is narrow,  a quadratic term in the phase $\Phi_{E}$ is negligible so that Fresnel's formula is reduced to a Fourier transform of the
shape of the aperture~(i.e., {\it Fraunhofer diffraction}). On the other hand, for $N_{F} \gtrsim 1$, Fresnel's formula is called a Fresnel transformation, and a resulting diffraction
pattern is a perfect shadow of the aperture~(i.e., {\it Fresnel diffraction}). By using the relation~(\ref{eq:corresp_2}), a corresponding quantity is also defined in the FHO case as:
\begin{equation}
\label{eq:fres_num}
N_{F}^{(\text{FHO})} \equiv \frac{\nu^{(1)}\Delta^{2}}{4},
\end{equation}
\noindent
and the relation~(\ref{eq:corresp_1}) is rewritten as
\begin{equation}
\label{eq:relation_nu_f_x}
\frac{4N_{F}^{(\text{FHO})}}{\nu^{(1)}\Delta}\left( \nu_{f} - \nu^{(0)} \right) \Longleftrightarrow \frac{4N_{F}}{2a}x.
\end{equation}
\noindent
As is the case of the single-slit diffraction, systems with the same value of $N_{F}^{(\text{FHO})}$ will have a response function $\tilde{h}(r)$ of equivalent properties.

\begin{figure}[t]
\includegraphics[clip,width=\linewidth]{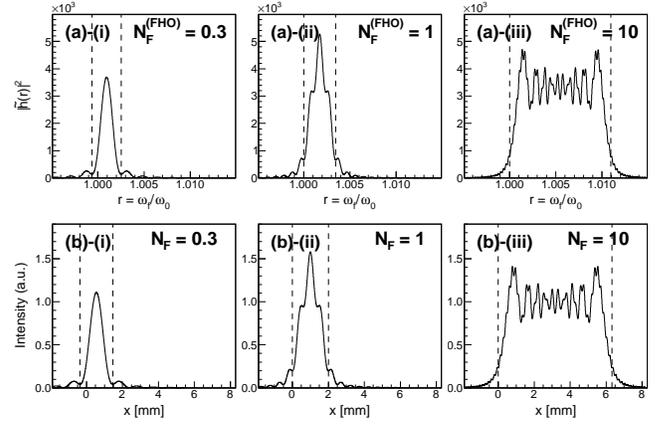}
\caption{\label{fig:diff_patt}(a) Frequency responses $\tilde{h}(r)$ for (i) $N_{F}^{(\text{FHO})}=0.3$, (ii) $N_{F}^{(\text{FHO})}=1$ and (iii) $N_{F}^{(\text{FHO})}=10$, with $\nu^{(0)}= 10~[\text{Hz}]$ and
$\nu^{(1)}=0.0003~[\text{sec}^{-2}]$. The ranges of resonant frequencies evaluated by Eq.~(\ref{eq:wid_omega_f}) are marked by dashed lines. (b) Diffraction patterns from a single slit
for (i) $N_{F}=0.3$, (ii) $N_{F}=1$ and (iii) $N_{F}=10$ with $\lambda = 1~[\mu\text{m}]$ and $r_{0}= 1~[\text{m}]$. Dashed lines represent the positions of $x = \pm \delta\bar{x}$.
For the definition of $\delta\bar{x}$, see the text.}
\end{figure}

Figures~\ref{fig:diff_patt}~(a) show the frequency responses $\tilde{h}(r)$ with different values of $N_{F}^{(\text{FHO})}$~(i.e., different values of $\Delta$). For reference, the intensity
patterns of single-slit diffraction with the same values of $N_{F}$~(i.e., corresponding values of $a$) are plotted in Figs.~\ref{fig:diff_patt}~(b). For both the phenomena, a
dramatic change of the frequency responses $\tilde{h}(r)$~(or the diffraction patterns) takes place around $N_{F}^{(\text{FHO})}\; (N_{F}) \approx 1$. Furthermore, the behavior of $\tilde{h}(r)$ on
$N_{F}^{(\text{FHO})}$ is in excellent agreement with that of the diffraction patterns on $N_{F}$. 

The observed correspondence between the FHO and the single-slit diffraction can be interpreted as follows: it is obvious from Eq.~(\ref{eq:res_fnc_simple}) that the FHO with
slowly-varying frequencies can be viewed as diffraction of waves in the {\it frequency} domain with time $t$ to be an independent variable, whereas the single-slit diffraction is 
discussed in the {\it space} domain. Thus, the frequency $\omega(t)$, moving in the {\it frequency} domain during a time window $\Delta$, is interpreted, in the case of single-slit diffraction,
as the incremental {\it space} coordinate $\xi$ on the slit from $0$ to $2a$, and the constant frequency $\omega_{f}$ as an observation point, i.e., the space coordinate $x$ on the
screen~[see the relation~(\ref{eq:relation_nu_f_x})].

A key feature common to both phenomena is a quadratic term in the phase functions of Eqs.~(\ref{eq:phase_FHO}) and (\ref{eq:phase_diff}), which yields Fresnel integrals.
In the FHO case, such a phase term comes from the difference of phase advance, i.e., the phase slippage between the oscillator and the driving force. In the single-slit diffraction case,
on the other hand, such a phase term comes from Fresnel's approximation of the optical path lengths of accumulated spherical waves.  We summarize the correspondence 
relations between the FHO and the single-slit diffraction in Table~\ref{tab:corresp}.

\begin{table}[t]
\begin{center}
\caption{\label{tab:corresp}%
Correspondence relations between the FHO and the single-slit diffraction.}
\begin{ruledtabular}
\begin{tabular}{cc}
FHO & Single-slit Diffraction \\ \hline
 $\omega_{f}$~(Driving force) & Position on screen \\
$\omega$~(Oscillator) & Position on slit \\ 
Variation of $\omega$ in $\Delta$, $2\pi\nu^{(1)}\Delta$ & Aperture size $D=2a$ \\
\begin{tabular}{c} Phase slippage between\\oscillator and force \end{tabular} & \begin{tabular}{c} Variation of optical \\ path length \end{tabular} \\
A quantity $N_{F}^{(\text{FHO})} \equiv \nu^{(1)}\Delta^{2}/4$ & Fresnel number $N_{F}=a^{2}/(\lambda r_{0})$ \\
\end{tabular}
\end{ruledtabular}
\end{center}
\end{table}

For quantitative discussion, we evaluate the range of resonant frequencies for the driving force, $2\delta \bar{\omega}_{f}$, using the analogies between the FHO and the single-slit diffraction. 
To clarify the situation, we start with the light diffraction case: for the Fraunhofer regime~($N_{F} \ll 1$), it is well known that the width $2\delta \bar{x}$ of a principal peak is
obtained from the slit-screen distance $r_{0}$ and the angle $\theta$, which defines a destructive phase relation between the wavelets from the both edges of the aperture, and is
given by $2\delta\bar{x} \approx 2 r_{0} \theta \approx \lambda r_{0} / (2a)$. For the Fresnel regime~($N_{F} \gtrsim 1$), the width of a rectangular pattern is almost the same as that of the
aperture, namely, $2\delta\bar{x} \approx 2a $. Now, the derivation of $\delta \bar{\omega}_{f}$ is straightforward: with the correspondence relations (\ref{eq:fres_num}) and (\ref{eq:relation_nu_f_x}),
we obtain:
\begin{equation}
\label{eq:wid_omega_f}
2\delta\bar{\omega}_{f} = 
\begin{cases}
2\pi/\Delta & \text{(for the Fraunhofer regime)} \\
2\pi\nu^{(1)}\Delta & \text{(for the Fresnel regime)}.
\end{cases}
\end{equation}
\noindent
The evaluated ranges for different values of $N_{F}^{(\text{FHO})}$ are indicated by dashed lines in Figs.~\ref{fig:diff_patt}~(a).  As we see from the figures, our evaluation is valid both for the 
Fraunhofer and Fresnel regimes. We notice that, for the FHO case, the center of resonant frequencies is given by $\bar{\omega}_{f} = 2\pi\nu^{(0)}+\pi\nu^{(1)}\Delta$.
 
As another example of the analogies between FHOs with time-varying frequencies and light diffraction, let us consider a HO with a time-varying natural frequency exposed continuously
to a sinusoidal force with a constant frequency. Here, we suppose that the frequency $\omega(t)$ of the oscillator varies linearly and coincides with the frequency $\omega_{f}$ of the
driving force at $t=0$.  In this case, a particular solution is obtained just by setting $r = 1$  and replacing $\Delta$ with $t$ in the response function of Eq.~(\ref{eq:res_fnc_simple}), namely:
\begin{equation}
\label{eq:res_fnc2}
\begin{split}
\tilde{h}(t) \simeq& \frac{e^{-i\phi_{0}}}{\sqrt{2\nu^{(1)}}} \displaystyle \int^{\sqrt{2\nu^{(1)}}t}_{0}\exp\left( i \displaystyle \frac{\pi}{2}\tilde{t}^{2} \right)d\tilde{t} \\
     =& \frac{e^{-i\phi_{0}}}{\sqrt{2\nu^{(1)}}} \left[ C\left(\sqrt{2\nu^{(1)}}t \right) + i S\left( \sqrt{2\nu^{(1)}}t \right) \right], 
\end{split}
\end{equation}
\noindent
which is in turn a function of $t$, and thus describes the time evolution of the oscillation amplitude, together with the damping factor $A(t)$~[see Eq.~(\ref{eq:envelop})]. The expression of Eq.~(\ref{eq:res_fnc2})
is quite similar to a diffraction formula for so-called knife-edge diffraction. 
In what follows, we neglect the damping factor $A(t)$, which does not stem from the presence of the driving force, in order to highlight the response of the oscillator and to compare it to knife-edge diffraction.   

\begin{figure}[t]
\includegraphics[clip,width=\linewidth]{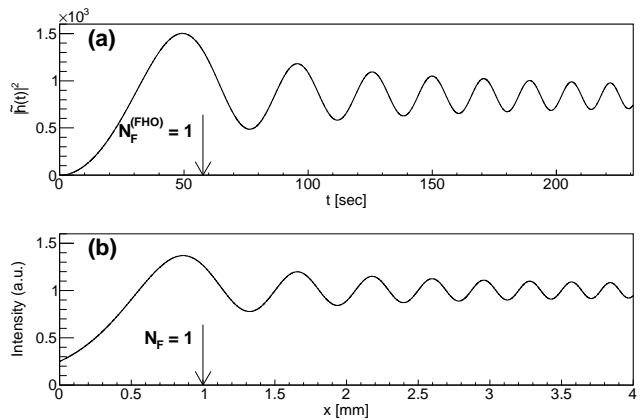}
\caption{\label{fig:evo_amp}(a) Time evolution of the squared amplitude $|\tilde{h}(r)|^{2}$ with $\nu^{(0)}= 10 ~[\text{Hz}]$ and $\nu^{(1)} = 0.0003 ~[\text{sec}^{-2}]$. An arrow indicates the time $t = \delta t$
corresponding to $N_{F}^{(\text{FHO})}=1$.  For the definition of $\delta t$, see the text. (b) Intensity pattern for light diffraction from a knife-edge obstacle with $\lambda = 1 ~[\mu\text{m}]$ and $r_{0} = 1 ~ [\text{m}]$. 
The obstacle is placed at $x \leq 0$. An arrow indicates the screen position $x$ corresponding to $N_{F}=1$.}
\end{figure}

Taking the limit $2a \rightarrow +\infty$ in Eq.~(\ref{eq:res_func_diff}) gives the expression of $\tilde{h}_{E}$ for knife-edge diffraction:
\begin{equation}
\label{eq:res_func_diff2}
\begin{split}
\tilde{h}_{E}(x) =& \int^{+\infty}_{0} \exp \left[ \frac{i\pi}{\lambda r_{0}} \left( \xi - x \right)^{2}  \right]d\xi \\
               =& \sqrt{\frac{\lambda r_{0}}{2}} \left[ \left\{ C\left( \sqrt{\frac{2}{\lambda r_{0}}}x \right) + \frac{1}{2}  \right\} \right. \\
                &+ \left. i \left\{ S\left( \sqrt{\frac{2}{\lambda r_{0}}}x \right) + \frac{1}{2}\right\}  \right].
\end{split}
\end{equation}
\noindent
Note that, by comparing the arguments of the Fresnel integrals in Eqs.~(\ref{eq:res_fnc2}) and (\ref{eq:res_func_diff2}), we can obtain a similar relation to Eq.~(\ref{eq:corresp_2}), namely:
\begin{equation}
\nu^{(1)}t_{\text{obs}}^{2} \Longleftrightarrow \frac{x_{\text{obs}}^{2}}{\lambda r_{0}},
\label{eq:rel_knife_edge}
\end{equation}
\noindent
where $t_{\text{obs}}$ and $x_{\text{obs}}$ are the observation time and position for the FHO and knife-edge diffraction cases, respectively.

Figure~\ref{fig:evo_amp}~(a) illustrates the time evolution of the squared oscillation amplitude~(or, equivalently, the energy of the oscillator), together with an intensity pattern for light
diffraction from a knife-edge obstacle~[Fig.~\ref{fig:evo_amp}~(b)]. We see that the time evolution of the oscillation energy behaves like a knife-edge diffraction pattern: that is, the energy
increases monotonically until $t \lesssim 60 ~[\text{sec}]$ and then exhibits small beating~(in other word, we could say that the oscillator is in a  quasi-stationary state). Asymptotically, it
approaches to~[see Eq.~(\ref{eq:res_fnc2})]:
\begin{equation}
|\tilde{h}(t)|^{2} \xrightarrow{t \rightarrow +\infty} \displaystyle \frac{1}{2\nu^{(1)}} \left[ C^{2}(+\infty) + S^{2}(+\infty) \right] = \displaystyle \frac{1}{4\nu^{(1)}}.
\label{eq:ene_limit}
\end{equation} 

The time duration $\delta t$ in which the driving force efficiently supplies kinetic energy to the oscillator is estimated by using the analogies: in knife-edge diffraction, a "good measure" of
the fringe width of diffraction  patterns, $\delta x$, is given by the condition that the corresponding Fresnel number, $N_{F} = \delta x^{2}/(\lambda r_{0})$, becomes unity~[see Fig.~\ref{fig:evo_amp}~(b)]. 
Similarly, from Eq.~(\ref{eq:rel_knife_edge}), we have the time duration $\delta t$:
\begin{equation}
\label{eq:time_dur}
\delta t = \frac{1}{\sqrt{\nu^{(1)}}} \approx 60 ~[\text{sec}], 
\end{equation}
\noindent 
with $\nu^{(1)} = 0.0003~[\text{sec}^{-2}]$.

\section{Summary}
In summary, we investigated a simple FHO with slowly-varying frequencies. We demonstrated that the time evolution of such a system can be written in a simplified form
using Fresnel integrals. As a result, we found that FHOs with slowly-varying frequencies can be viewed as diffraction of waves in the frequency domain, and therefore
are equivalent to diffraction of light. Also we showed two examples to see the similarities between the two phenomena, and derived simple formulae for the quantities
which characterize the systems. We expect that our formulation as well as such simple formulae is applied to, e.g., accelerator physics and provides a simple and intuitive
approach to the phenomenon of "resonance crossing", which is a central issue in a ring-type particle accelerator design~\cite{cross_res,cross_res2}. As a matter of fact, we applied
our formulation to the design of an aborted-beam-handling system for a new synchrotron light source accelerator~\cite{abort_beam}.
In this system, a sinusoidal force is applied to aborted beams, whose betatron frequency varies with time due to energy loss by synchrotron radiation. A proper choice of frequency of
the sinusoidal force is essential to enlarge the amplitude of betatron oscillation and to reduce the beam density.
Our findings will be also applicable to plasma physics, where the problem of passage through resonance with slowly-varying parameters is of great importance~\cite{cross_res3}.


\vspace{\baselineskip}

\appendix
\section{\label{app:green_fnc}Derivation of Green's function}
In this appendix, we present the derivation of the Green's function of Eq.~(\ref{eq:green_fnc}). With the aid of the method of "variation of constants", the Green's function of an inhomogeneous
differential equation such as Eq.~(\ref{eq:basic_eq}) is in general written in the form:
\begin{equation}
\label{eq:general_green_fnc}
\begin{split}
G(t, t^{\prime}) &= \frac{\left| \begin{array}{cc}
                                     x_{1}(t^{\prime}) & x_{2}(t^{\prime}) \\
                                     x_{1}(t)            & x_{2}(t)
                                     \end{array}   \right|}{W(x_{1}, x_{2})(t^{\prime})} \\
                   &\equiv \frac{ x_{1}(t^{\prime})x_{2}(t) - x_{1}(t)x_{2}(t^{\prime})}{x_{1}(t^{\prime})\dot{x}_{2}(t^{\prime}) - \dot{x}_{1}(t^{\prime})x_{2}(t^{\prime})},                 
\end{split}
\end{equation}
\noindent
where $x_{1}$ and $x_{2}$ are independent solutions for the corresponding homogeneous differential equation, and $W$ is the Wronskian. Thus, in our case, the problem comes down to solving
the following homogeneous equation:
\begin{equation}
\ddot{x} + \omega^{2}(t)x = 0.
\label{eq:homo_eq}
\end{equation}

To solve the above equation, we use the so-called eikonal approximation~\cite{eikonal}; that is, it is assumed that a solution of Eq.~(\ref{eq:homo_eq}) is of the form:
\begin{equation}
x(t) = a(t)e^{i\varphi(t)},
\label{eq:sol_eikonal}
\end{equation}
where the envelope function $a(t)$ varies very slowly compared to oscillation of $x(t)$, namely:
\begin{equation}
\frac{\ddot{a}(t)}{a(t)} \ll \omega^{2}(t).
\label{eq:cond_eikonal}
\end{equation}

Substituting Eq.~(\ref{eq:sol_eikonal}) in Eq.~(\ref{eq:homo_eq}) and using the condition~(\ref{eq:cond_eikonal}), we obtain:
\begin{align}
\left[ \dot{\varphi}(t) \right]^{2} =& \omega(t)^{2}, \label{eq:eq_eikonal_1} \\
2\dot{a}(t)\dot{\varphi}(t) + a(t)\ddot{\varphi}(t) =& 0. \label{eq:eq_eikonal_2}
\end{align}

It follows from Eq.~(\ref{eq:eq_eikonal_1}) that:
\begin{equation}
\varphi(t) = \pm \left[ \int^{t}_{0}\omega(\tau)d\tau + \varphi_{0} \right],
\label{eq:sol_phase}
\end{equation}
\noindent
where $\varphi_{0}$ is an integration constant.

The substitution of Eq.~(\ref{eq:sol_phase}) into Eq.~(\ref{eq:eq_eikonal_2}) gives:
\begin{equation}
\frac{d}{dt}\left[ \ln a(t) + \frac{1}{2}\ln\omega(t) \right] = 0,
\label{eq:eq_eikonal_2_mod}
\end{equation}
\noindent
and we have:
\begin{equation}
a(t) = \frac{\alpha}{\sqrt{\omega(t)}},
\end{equation}
\noindent
where $\alpha$ is a constant.

Thus, two independent solution of Eq.(\ref{eq:sol_eikonal}) are given by:
\begin{equation}
x_{1,2}(t) = \frac{\alpha}{\sqrt{\omega(t)}}\exp\left[ \pm i \left( \int^{t}_{0}\omega(\tau)d\tau + \varphi_{0} \right) \right],
\label{eq:sol_homo_eq}
\end{equation}
\noindent
and the substitution of Eq.~(\ref{eq:sol_homo_eq}) into Eq.~(\ref{eq:general_green_fnc}) yields the Green's function of Eq.~(\ref{eq:green_fnc}). Note that the condition~(\ref{eq:cond_eikonal}) is 
clearly fulfilled under the assumption~(\ref{eq:assump}).

\bibliography{pre_draft}

\end{document}